\documentclass[]{spie}  
\usepackage[dvips]{graphicx}

\title{Atmospheric time constant with MASS and FADE}

\author{Andrei Tokovinin 
\skiplinehalf
Cerro Tololo Inter-American Observatory, Casilla 603, La Serena, Chile
}

\authorinfo{E-mail:   atokovinin@ctio.noao.edu }

\begin{document} 
  \maketitle 

\begin{abstract}

The   approximate  nature   of  the   adaptive-optics   time  constant
measurements  with MASS  is examined.   The calibration  coefficient C
derived from numerical simulations of polychromatic scintillation shows
dependence  on the  height of  the turbulence  layer, wind  speed, and
seeing. The previously recommended value  of C=1.27 is a good match to
typical  conditions,  while  C can  vary  from  0.6  to 1.6  in  other
circumstances. For two  nights, MASS  was  compared with the
time  constant  measured  with   adaptive  optics,  and  the  expected
agreement was found. We  show that the single-layer approximation used
in  some AO  systems to  derive the  AO time  constant can  give wrong
results. A better  approach is to estimate it from  the speed of focus
variation  (the   FADE  method).   The   analysis  of  the   speed  of
scintillation  developed recently  by  V.~Kornilov will  lead to  more
accurate measurements of the AO time constant with MASS.
\end{abstract}


\keywords{Site testing; MASS; Adaptive Optics}

\section{Definitions and context}
\label{sect:definition}  

The adaptive optics atmospheric  time constant $\tau_0$ depends on the
wind speed and turbulence profiles.  It is defined as
\begin{equation}
\tau_0 = 0.314 r_0/\overline{V}_{5/3} =
0.057 \; \lambda_0^{6/5} \left[ \int_o^{\infty} C_n^2(h) \; V^{5/3}(h)\; {\rm
    d} h \right] ^{-3/5} ,
\label{eq:tau0}
\end{equation}
where  $C_n^2(h)$ is  the  vertical profile  of  the refractive  index
structure constant, $V(h)$  is the vertical profile of  the modulus of
the wind  speed, $h$ is the  altitude above site. In  the following we
always  assume that  $\tau_0$ refers  to the  wavelength  $\lambda_0 =
0.5$~$\mu$m and  observations at zenith. As the correction  for the zenith
distance depends on the unknown wind direction, it is simply ignored
\cite{Travouillon}.

The  MASS instrument  implements an  approximate method  \cite{tau} of
estimating $\tau_0$ from  the {\it differential-exposure scintillation
  index}, DESI. The  $\sigma^2_{DESI}$ is computed  for the smallest
2-cm  MASS aperture  as a  differential  index between  1~ms and  3~ms
exposures.  A formula
\begin{equation}
\tau_{MASS}  = 0.175\,{\rm ms}\; (\sigma^2_{DESI} )^{-0.6}
\label{eq:tmass}
\end{equation}
has been  suggested on  the basis of  limited data on  real turbulence
profiles. This formula is implemented in the standard MASS data processing

We found  later by means of simulations  that $\tau_{MASS}$ calculated
from (\ref{eq:tmass})  needs a corrective  coefficient around $C=1.27$
The true time constant could be derived from the MASS data by applying
this correction and  adding the contribution of the  ground layer (GL)
which is not sensed by MASS (but measured with MASS-DIMM).  Therefore,
the AO time constant is estimated by MASS-DIMM as
\begin{equation}
\tau_0^{-5/3} = (1.27\,\tau_{MASS})^{-5/3} + 
(0.057)^{-5/3} \lambda_0^{-2} \; V_{GL}^{5/3} \; (C_n^2 {\rm d} h)_{GL}.
\label{eq:recipe}
\end{equation}
The intrinsic accuracy of such estimate was evaluated to be $\pm 20$\%
or better.

Travouillon  et  al.    \cite{Travouillon}  have  derived  a  somewhat
different correction factor $C=1.73$  by calculating $\tau_0$ from the
turbulence profile  measured by MASS  and using NCEP  wind velocities.
Even   larger  factors   of   2.45  and   2.11   were  determined   in
Ref.~\citenum{Tok06}   by   the   same   method.   This   prompted   a
re-investigation of this matter by doing new simulations and comparing
with alternative techniques.

\section{Numerical simulations}
\label{sec:simul}

We simulate one  phase screen at a given height  $h$ and calculate the
intensity at the ground by  means of the program {\tt simatmpoly.pro}.
The intensity pattern is a  weighted sum for several wavelengths. Here
we  approximate the  spectral response  of  MASS by  4 wavelengths  of
[400,450,500,550]\,nm  with weights [0.31,  0.885, 0.60,  0.27].  This
should mimic  the response of  TMT MASS-DIMMs, as studied  by Kornilov
\cite{Kornilov2006}.   He   found  for  these   instruments  effective
wavelength  474\,nm  and the  bandwidth  99\,nm  (response curve  {\tt
  without.crv}).   For our  4-wavelength approximation,  the effective
wavelength and  bandwidth are  470\,nm and 116\,nm,  respectively. The
simulation  does not  include  the light-source  spectrum, assuming  it
flat.

\begin{figure}[ht]           
\center          
\includegraphics[width=8cm]{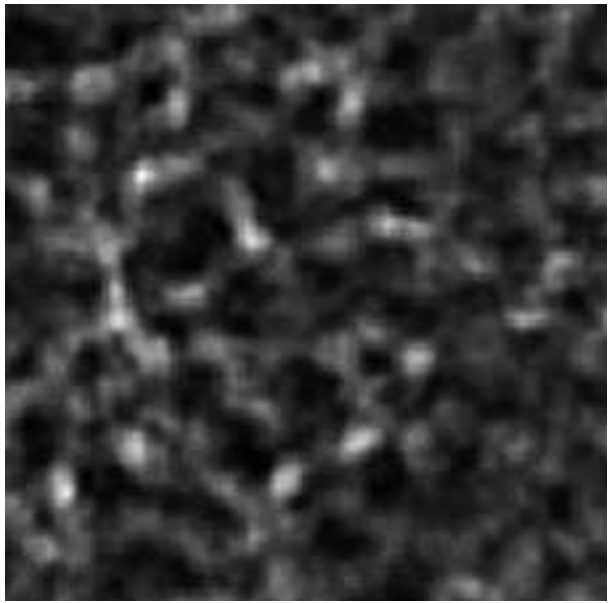} 
\caption{Fragment of the simulated scintillation pattern produced on
  the ground by a turbulent layer at 10\,km with $0.5''$ seeing. The
  scintillation index is 0.27, the size of the fragment is 0.5\,m. 
\label{fig:intens}   }
\end{figure}

The  simulated  intensity  distributions  (Fig.~\ref{fig:intens})  are
saved in a binary file. In  the previous simulator they were used in a
Monte-Carlo  approach where  the  2-cm MASS  aperture was  ``dragged''
through the  intensity screen, while  1-ms and 3-ms exposure  time was
emulated by suitable blurring of  the aperture in one direction and by
3x binning.   Here we take a  more direct approach  and calculate DESI
with the spatial filter
\begin{equation}
P(f_x) = {\rm sinc}(f_x b) -  {\rm sinc}(3 f_x b),  
\label{eq:Pfilt}
\end{equation}
where $f_x$ is the component of  spatial frequency, $b = V t_{exp}$ is
the  blur in  the  x-direction caused  by  the wind  speed $V$  during
exposure time $t_{exp} = 1$\,ms and ${\rm sinc} (x) = \sin(\pi x)/(\pi
x)$.  The circular aperture of diameter $d$ implies the filter
\begin{equation}
A(f) = \frac{2J_1(\pi f d)}{\pi f d}. 
\label{eq:Afilt}
\end{equation}

The energy  spectrum of  the intensity is  multiplied by  the combined
filter  $(A P)^2$  and summed  over all  frequencies to  get  the DESI
index. By omitting all filters, we obtain the raw scintillation index,
by  using only  filter  $A$ --  the  scintillation index  in the  2-cm
aperture.   Results  for  a  test  case were  compared  with  previous
simulations and found to be in agreement, validating the code.

Considering that the calculation of intensity distribution is the most
time-consuming  task, we  simulate the  intensity screen  for  a given
seeing  and   propagation  distance   and  then  calculate   DESI  and
$\tau_{MASS}$ for a set of 12 wind speeds, from 10\,m/s to 65\,m/s, by
changing  only the  filter $P$.   The  calculation is  repeated for  3
distances to the layer, 5, 10,  and 15\,km, and for 5 values of seeing
$\varepsilon$,  from $0.3''$ to  $1.5''$.  Therefore  we cover  a wide
range of conditions, with a varying degree of saturated scintillation.
The  largest scintillation  index  is 0.94  (layer  at 15\,km,  seeing
$1.5''$).   In each  case, the  true time  constant at  $\lambda_0$ is
known,  $\tau_0  =  0.31  r_0/V  =  0.31  (0.101/\varepsilon)/V$.   We
determine the  corrective coefficient $C  = \tau_0 /  \tau_{MASS}$ for
each case.

\begin{figure}[ht]           
\center          
\includegraphics[width=8.5cm]{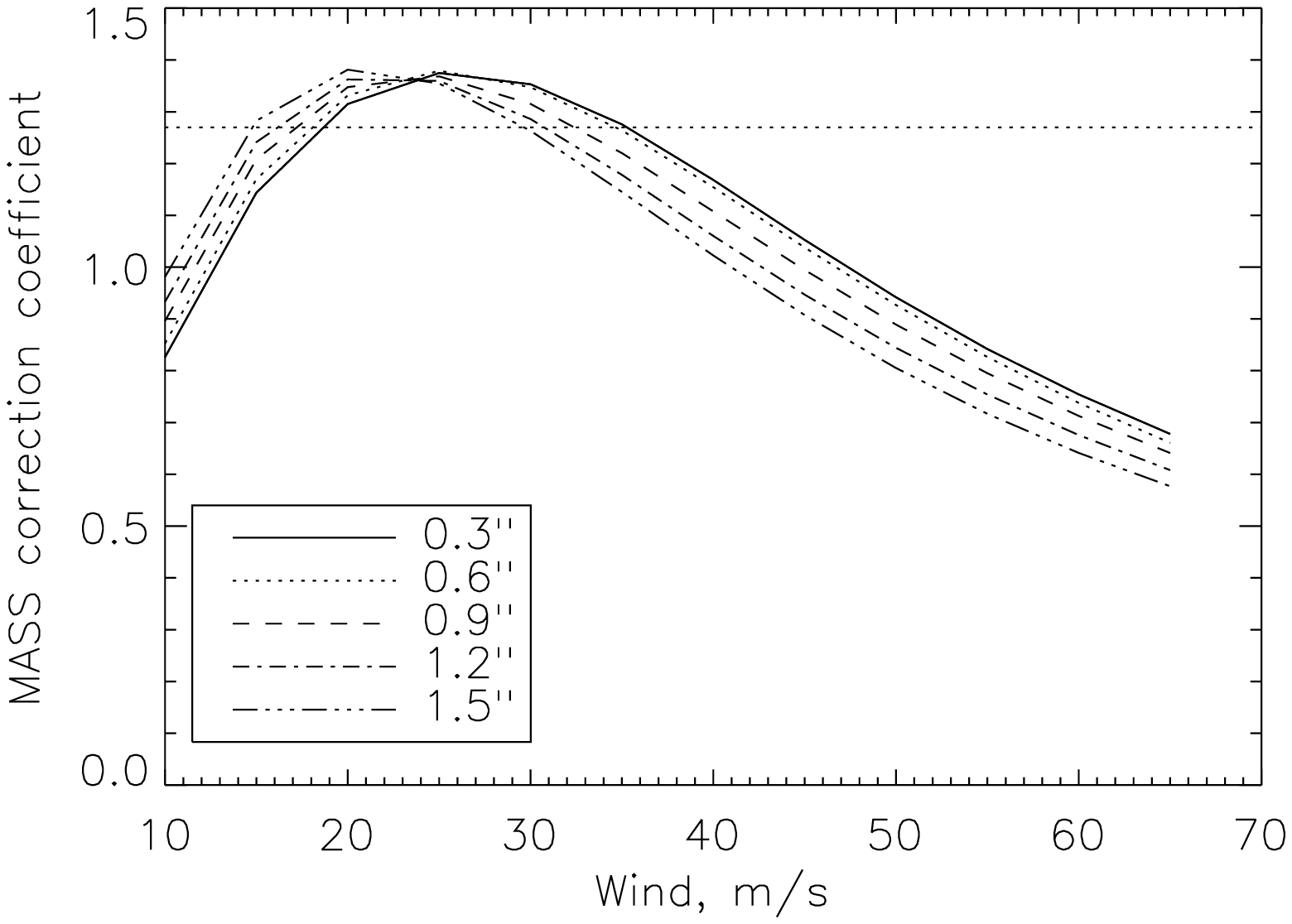} 
\includegraphics[width=8.5cm]{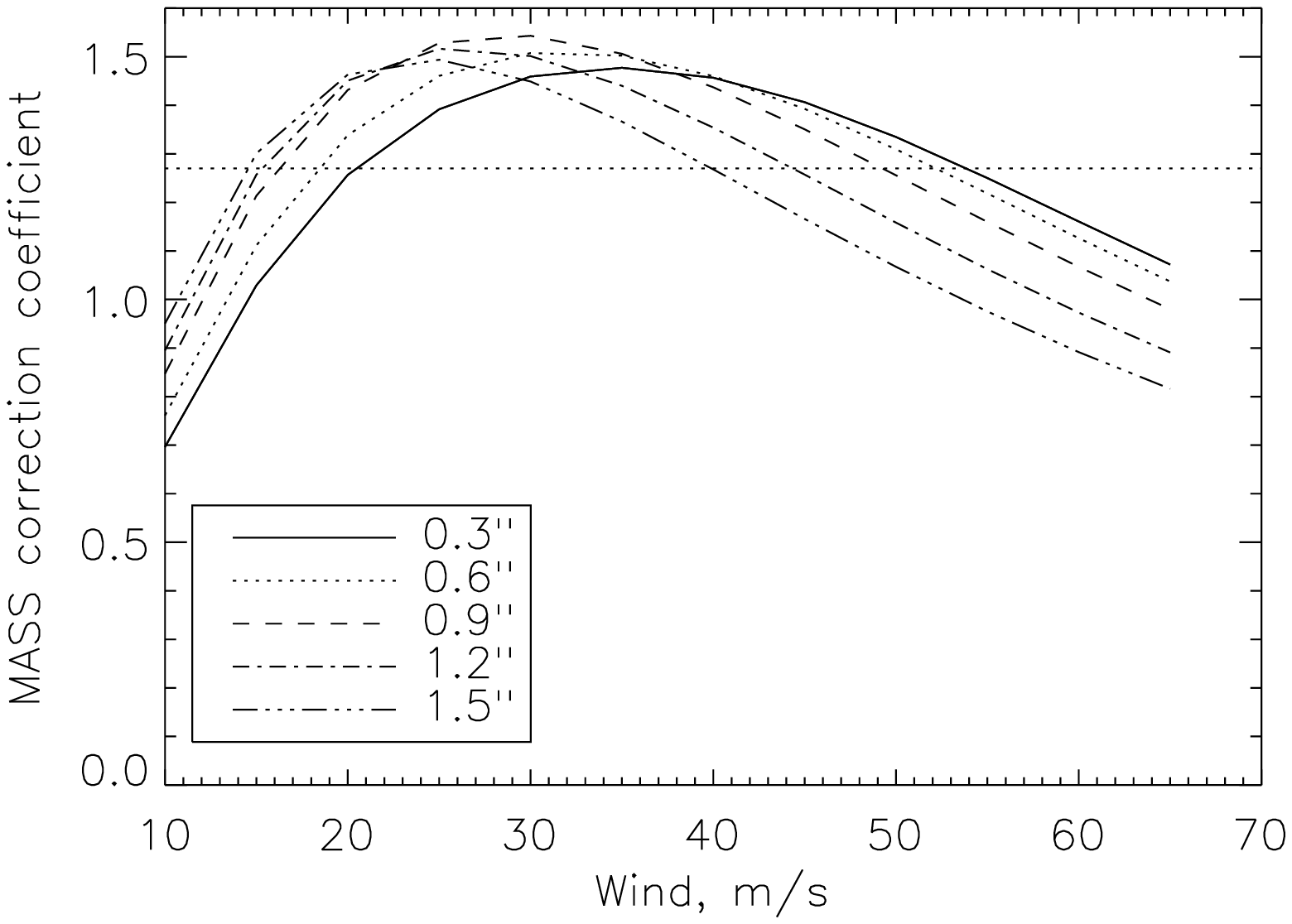} 
\caption{ Correction factor $C$ as a  function of the wind speed for a
  single  turbulent   layer  at  distance  5\,km   (left)  and  10\,km
  (right).  The turbulence  strength  of the  layer  (seeing) is  from
  $0.3''$ to $1.5''$. The dotted horizontal line shows $C=1.27$.
\label{fig:5km} }  
\end{figure}

\begin{figure}[ht]           
\center          
\includegraphics[width=8.5cm]{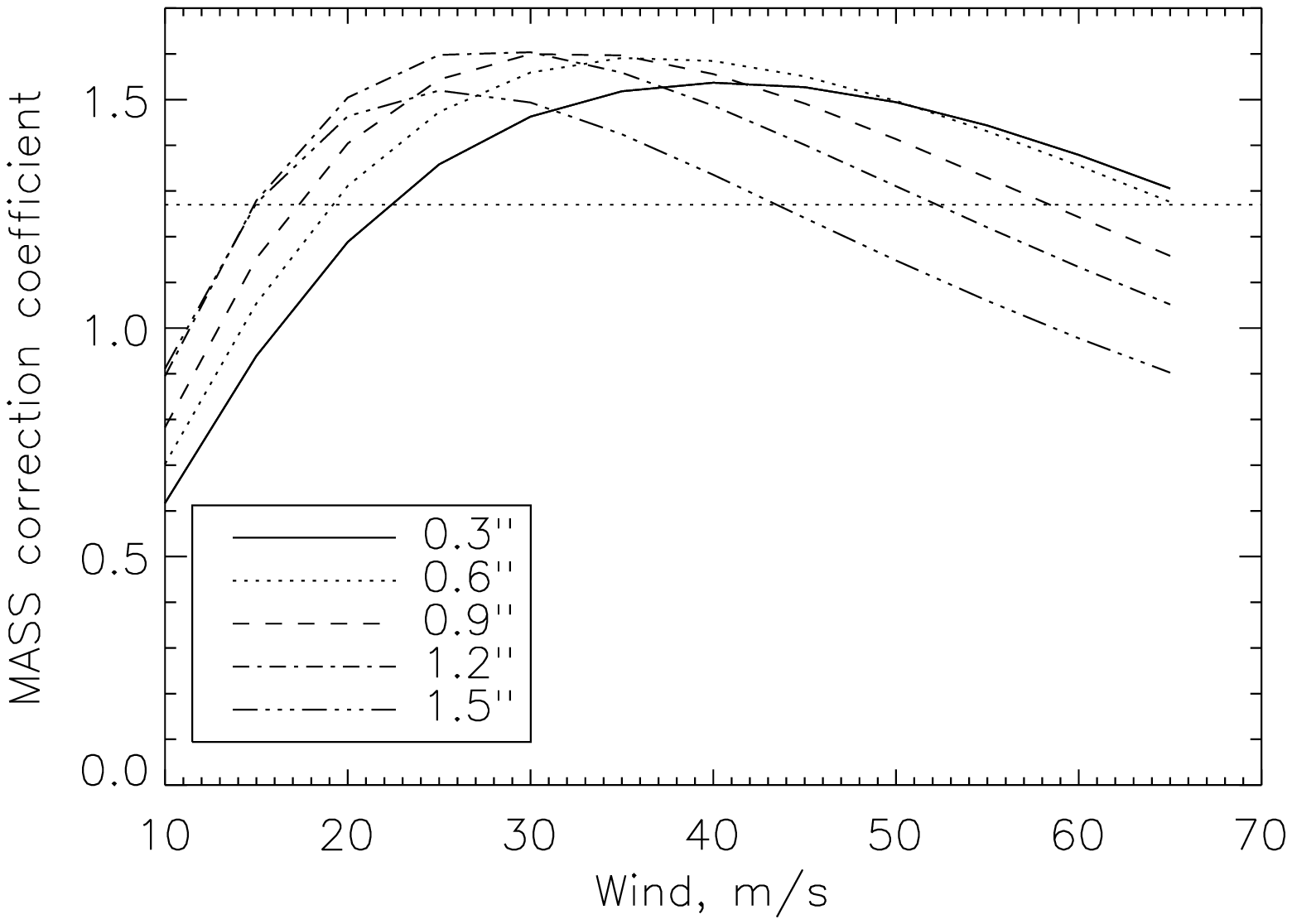} 
\includegraphics[width=8.5cm]{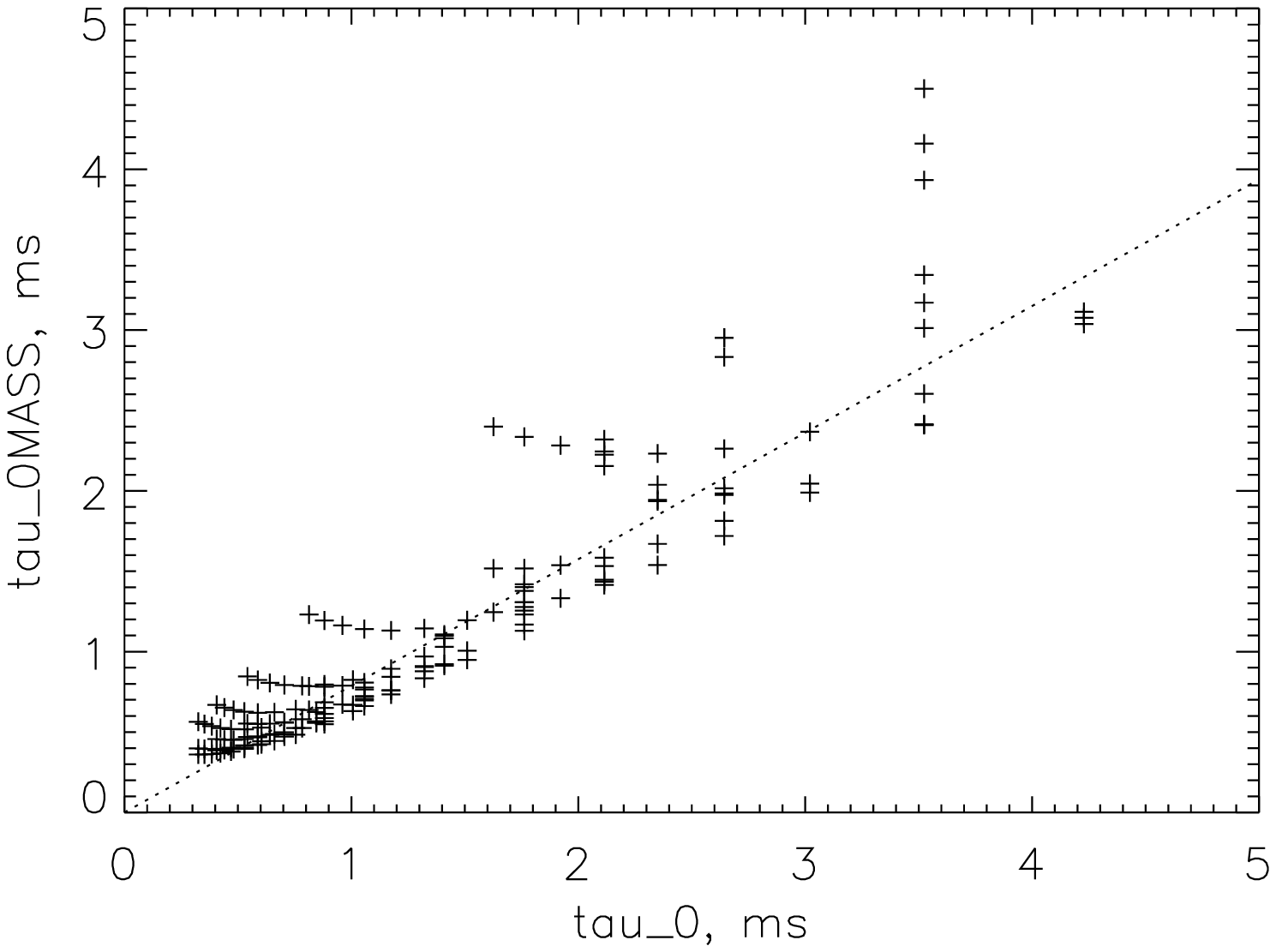} 
\caption{Left:   Correction factor $C$ as a  function of the wind speed for a
  single  turbulent   layer  at  distance  15\,km.
Right: all values of $\tau_{MASS}$ plotted against $\tau_{0}$, the
  dotted line corresponds to   $C = 1.27$. 
\label{fig:all} }  
\end{figure}

This  study clearly  shows that  estimates of  $\tau_0$ from  DESI are
approximate  and  can be  off  by  as much  as  two  times. No  single
corrective  coefficient $C$  can be  derived,  it varies  from 0.6  to
1.6.  For  typical conditions  ($V=30$\,m/s,  turbulence at  200\,mb),
$C=1.27$ appears to be a good choice.

\section{Comparison of MASS with FADE} 
\label{sec:FADE}

Any working AO system provides real-time data on turbulence from which
the AO  time constant can be  derived. For example,  the covariance of
atmospheric  defocus falls  to  1/2 of  its  maximum for  a time  lag
$t_{0.5} = 0.30 D/V$. It follows that $\tau_0 = 1.05 t_{0.5} (r_0/D)$.
A method of estimating  $\tau_0$ from half-time correlation of Zernike
aberrations  has  been proposed  by  Fusco  et  al.  \cite{Fusco04}  and
implemented in the NAOS at Paranal. It is valid for a single turbulent
layer, but the authors argue  that for several layers some ``average''
wind speed will result from  $t_{0.5}$. This is not quite true, as
we will see in a moment.

Theory  \cite{KT07,TKF08} shows that  the temporal  structure function
(SF)  of  defocus  $D_4(t)$  produced  by  a  combination  of  several
atmospheric layers has the form
\begin{equation}
D_4(t) \approx 1.94 D^{5/3} \sum_i r_{0,i}^{-5/3} K_4(2 V_i t/D ,
\epsilon) + N , 
\label{eq:mod}
\end{equation}
where  $D$  is  the  telescope  diameter, $\epsilon$  is  the  central
obscuration ratio,  $r_{0,i}$ and $V_i$  are the Fried  parameters and
wind speeds of  the layers, $N$ is the term  caused by the measurement
noise.  The  function $K(x,y)$ is defined  in Ref.~\citenum{TKF08}. It
has a  quadratic initial  part at $t  \ll t_{0.5}$, where  small focus
changes  are  proportional  to  time.  In  this  {\em  short-exposure}
regime, the speed  of defocus variation is related  to the integral of
$V^2(h)  C_n^2(h) {\rm  d}h$  -- the  2-nd  moment of  the wind  speed
$\overline{V}_2$.    It  is   rather   close  to   the  5/3rd   moment
$\overline{V}_{5/3}$ used  in the definition of  $\tau_0$.  The method
to estimate $\tau_0$ from the  speed of focus variation is called FADE
(FAst DEfocus) \cite{KT07,TKF08}. It is resumed by a simple formula
\begin{equation}
D_4(t) \approx 0.036 (D/r_0)^{-1/3} (t/\tau_0)^2 .
\label{eq:FADE}
\end{equation}

\begin{figure}[ht]           
\centerline{          
\includegraphics[width=8.5cm]{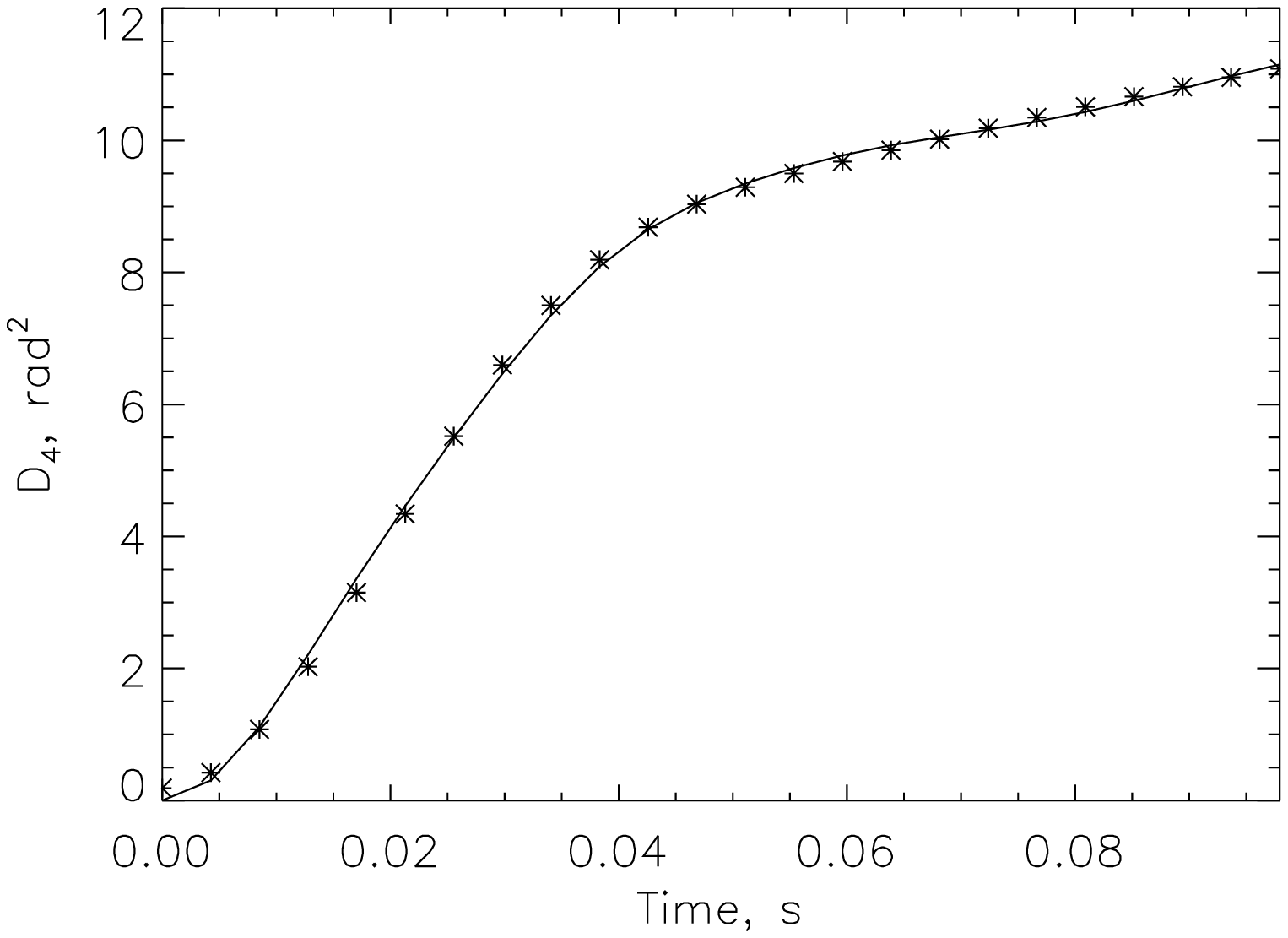} 
\includegraphics[width=8.5cm]{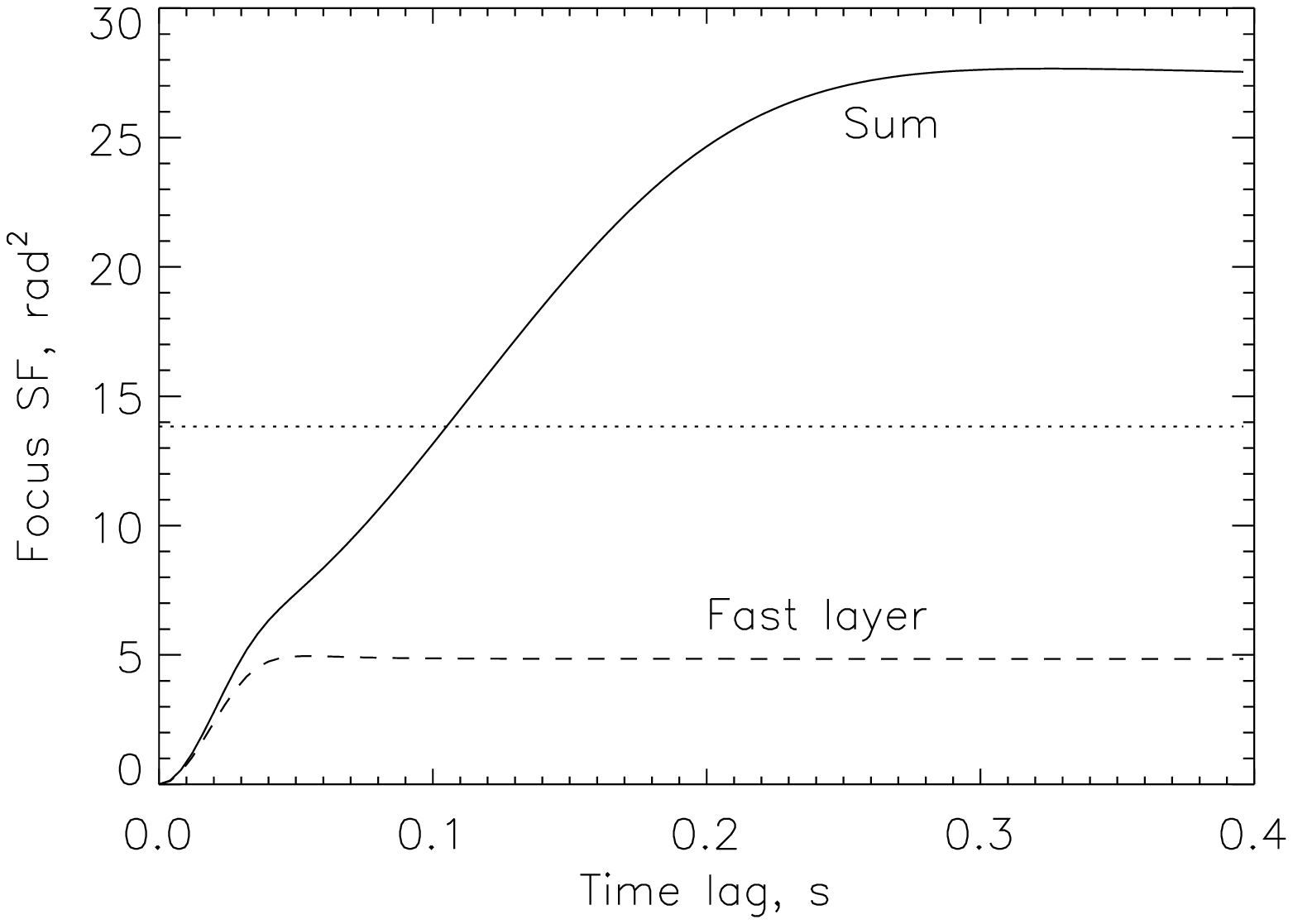} 
}
\caption{Left: example of the measured defocus SF (line) and its model
  by two layers  (asterisks). The vertical axis in  rad$^2$ at 500\,nm
  wavelength. 
Right:  SF of defocus  for two  layers with  $1''$ and  $0.4''$ seeing
moving  at  10  and  60\,m/s,  respectively.  the  dotted  line  shows
half-saturation and corresponds to $t_{0.5}$.
\label{fig:SF} }  
\end{figure}

Figure~\ref{fig:SF}  shows the SF  of defocus  measured with  the SOAR
Adaptive  Module (SAM)  on  October 2009.   In  this case  $D=4.1$\,m,
$\epsilon  = 0.24$.   Taking the  wind speed  of 50\,m/s,  we estimate
$t_{0.5}  \sim 25$\,ms,  or  6  loop cycles  of  SAM.  Therefore,  the
temporal sampling of  SAM (4.2\,ms) is adequate for  applying the FADE
technique.  By  fitting the  initial part of  the SF with  a two-layer
model (\ref{eq:mod}), we derive the defocus speed and $\tau_0$.

A simple  test case presented  in Fig.~\ref{fig:SF}, right,  shows why
the half-time  method can give wrong  results.  The $t_{0.5}=0.105$\,s
is  essentially determined here  by the  strongest and  slowest layer,
leading to the estimate $\tau_0  = 2.39$\,ms. The actual time constant
$\tau_0  = 1.15$\,ms is  mostly produced  by the  weak and  fast layer
which makes the  dominant contribution to the speed  of defocus and to
the  shape  of the  initial  (quadratic) SF.   The  error of  the
half-time method to estimate $\tau_0$ in this case is 2.1 times.

So  far, the SAM  data were  collected and  processed for  two nights,
August 31 and October 2, 2009 (hereafter nights 1 and 2). AO loop data
were  recorded   several  times  during   $\sim  12$\,s.   Atmospheric
variations of defocus  were derived from the signals  of the corrector
(DM),  accounting  for  the   frequency  response  of  the  closed  AO
loop. Voltages  are transformed into Zernike  coefficients in radians
at $\lambda_0$.   The variance of the  low-order Zernike coefficients
is used to estimate the Fried parameter $r_0$ (seeing).

Derivation   of   $\tau_0$   from   the   loop   data   involves   few
subtleties. First,  we suppress  all frequencies above  45\,Hz because
the data show  focus vibration at 46\,Hz (this  low-pass filtering has
little effect  on the results, however). Second,  the estimates depend
on the  maximum time  lag used in  the model-fitting. This  length was
varied from just 4 first points to 0.1\,s and 0.4\,s.  With increasing
lag,   the   estimates   $\tau_{0,SAM}$   also  become   larger,   see
Table~\ref{tab:fade}. The fitted models show that on night 1 there was
a fast-moving layer.

\begin{table}[ht]
\center
\caption{Average parameters from the FADE model fits}
\label{tab:fade}
\medskip
\begin{tabular}{l | ccc | cccc }
\hline
Parameter & \multicolumn{3}{c|}{Night 1} & \multicolumn{3}{c}{Night 2} \\
          & 4pt & 0.1s & 0.3s & 4pt & 0.1s & 0.3s \\
\hline
$\tau_{0,MASS}/\tau_{0,SAM}$ & 0.73 & 0.83 & 0.91 &   1.42 & 1.25 & 1.02 \\
$\overline{V}$, m/s                 & 30.5 & 36.6 & 27.6 & 24.6 & 27.8 & 19.0  \\
$V_{layer 2}$, m/s                 & - & 63.9 & 47.1    & -    & 64.9 & 32.8  \\ 
\hline
\end{tabular}
\end{table}

\begin{figure}[ht]           
\center         
\vspace*{0.5cm} 
\includegraphics[width=9cm]{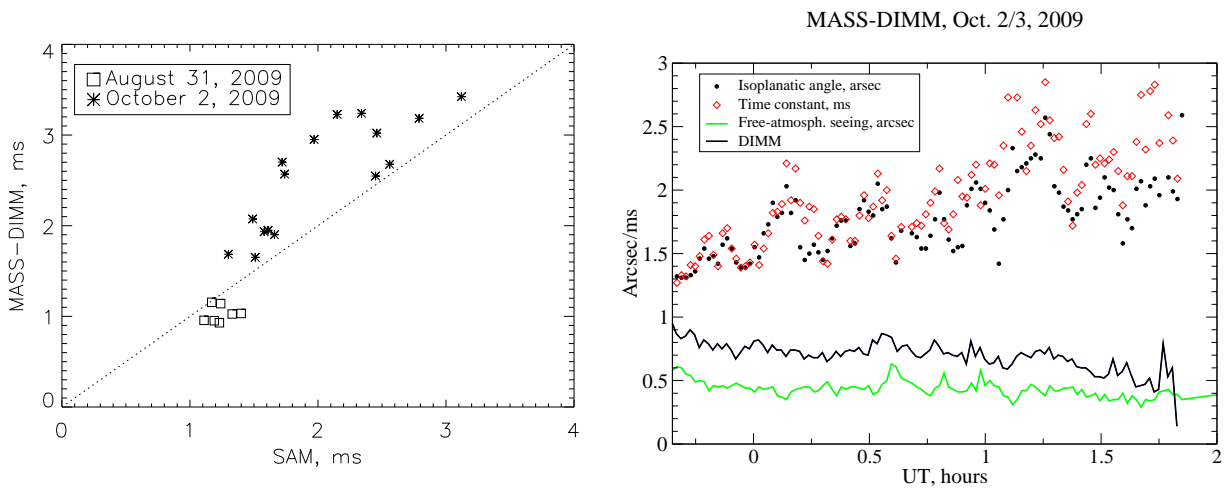} 
\includegraphics[width=8.0cm]{091002mass.eps} 
\vspace*{0.5cm} 
\caption{Left: Comparison of  simultaneous $\tau_0$ from FADE (fitting
  0.1\,s) and MASS; the dashed line shows equality.
Right: Summary plot of atmospheric  conditions on October 2, 2009 from
the Cerro Pach\'on MASS-DIMM site monitor.
\label{fig:FADE}   }
\end{figure}

The MASS-DIMM  monitor at Cerro Pach\'on provided  data for comparison
with  FADE. We  used the  ground wind  speed and  applied  the formula
(\ref{eq:recipe}) with $C=1.27$ to  get $\tau_{0,MASS}$. The seeing on
the night  1 was  fast and bad,  while the  conditions on the  night 2
(Fig.~\ref{fig:FADE}, right) were closer  to typical, but only 4 hours
were  clear.  Different structure  of turbulence  on these  two nights
leads    to   different    systematics   between    MASS    and   FADE
(Fig.~\ref{fig:FADE}, left).   We cannot tell whether  $C=1.27$ has to
be increased or decreased, it is ``about right''.

\section{Conclusions and further work} 
\label{sec:concl}

This  study reveals, once  again, the  approximate nature  of $\tau_0$
measurements from DESI implemented  in the MASS software. Depending on
the turbulence and  wind profiles, the bias of  MASS $\tau_0$ with the
recommended  correction  factor $C=1.27$  can  be  either positive  or
negative.   Attempts  to find  the  best  calibration coefficient  are
hampered  by  the   above-mentioned  dependence  on  conditions.   Our
simulations  show  that  the   correction  factor  $C=1.73$  found  by
Travouillon  et al.   \cite{Travouillon}  is too  high.  However,  the
simulations do not  account for the spectrum of the  star; if it moves
the  effective spectral  response  of MASS  blue-wards, the  correction
factor will be larger.

Despite  obvious shortcomings  of the  DESI  method, it  has a  strong
appeal, being  a simple and  no-cost addition to the  MASS instrument.
Considerable data on  $\tau_0$ have been accumulated to  date for many
sites worldwide.  We can do a better job on $\tau_0$ with MASS. Recent
analysis  by   Kornilov  \cite{Kornilov11}  uses   the  short-exposure
approximation, where the dependence  on wind is quadratic and separates
from  the spatial  dependence.  This  permits to  find  optimum linear
combinations  of  signals  from  MASS  apertures  and  their  pairwise
covariances to cancel the height dependence, similarly to what is done
for the  free-atmosphere seeing  and isoplanatic angle.   By comparing
this  new method with  DESI for  one site,  Kornilov finds  $C \approx
1.7$.

The second important improvement  consists in using a longer effective
exposure time when  the turbulence is slow. This  is possible with the
new  MASS  software which  records  relevant statistical  information.
Unfortunately, this information is lost in the archival MASS data that
cannot be cured post-factum; on  good nights with slow turbulence when
the DESI  signal is weak,  the $\tau_0$ derived  from it is  noisy and
biased \cite{Travouillon}.

Finally, it  becomes clear  that for all  turbulence-dependent optical
parameters (scintillation, Zernike  aberrations, and even differential
image motion) the speed of their temporal variation is directly related
to the wind-speed  2nd moment $\overline{V}_2$, and hence  can be used
to measure  $\tau_0$. It is  a practical matter  to choose one  or the
other optical tracer for which  the speed of variation can be measured
accurately. The  advantage of defocus in  this respect is  that, for a
given  aperture,  it produces  the  strongest  signal while  remaining
isotropic and immune to the instrument shake.  But a DIMM with two or,
preferably,  4  apertures  could  also  be a  promising  solution  for
measuring $\tau_0$, if a suitable theoretical analysis of its response
is done.

\acknowledgments 

The  work   on  $\tau_0$  has  greatly   benefited  from  stimulating
interaction    with    my    colleagues    V.~Kornilov,    M.~Sarazin,
T.~Travouillon, A.~Kellerer and others. 




\end{document}